\documentstyle[preprint,aps]{revtex}
\begin{document}
\preprint{UM-P-94/73, RCHEP-94/21}
\draft
\title{NON-ZERO ELECTRIC CHARGE OF THE NEUTRINO AND THE
SOLAR NEUTRINO PROBLEM}
\author{A.Yu.Ignatiev\cite{byline1} and
G.C.Joshi\cite{byline2}}
\address{Research Centre for High Energy Physics, School of
Physics, University of Melbourne, Parkville, 3052, Victoria,
Australia}
\maketitle
\begin{abstract}
It has recently been shown that the neutrino can
have non-zero electric charge in a number of gauge theories,
including the Minimal Standard Model.
Assuming non-zero neutrino charge, we
develop a new approach to the solar neutrino problem. The key idea is
that the charged neutrinos will be deflected by the Lorentz force while
they are crossing the solar magnetic fields. Such a deflection will
result in the anisotropy of the solar neutrino flux. Because of this
anisotropy, the solar
neutrino flux
registered on earth can be reduced as compared to the
Standard Solar Model
prediction. The mechanism is purely classical
and does not require neutrino oscillations, spin-flip or
neutrino decay. We discuss qualitatively the consequences of our
scenario for present and future solar neutrino experiments
as well as differences between our mechanism and other
proposed solutions.
\end{abstract}
\pacs{}

\section{introduction}
A long-standing puzzle at the crossroads of elementary
particle physics and astrophysics is a
deficiency
of the flux of solar neutrinos registered in four different
experiments. The data of the two-decade-long $Cl-Ar$
experiment
at Homestake \cite{home} have more recently been
complemented by the direct observation of solar
neutrinos at the Kamiokande water Cherenkov detector
\cite{kamioka} as well as by two different
$Ga-Ge$ experiments, SAGE \cite{sage} and GALLEX
\cite{gallex}. Being sensitive to
different parts of the solar neutrino spectrum, these four
experiments together provide the data which are difficult to
understand with a single set of solar parameters chosen
within the Standard Solar Models \cite{ssm1,ssm2}. Chances
are that we have to modify some aspects of neutrino behavior
in order to explain the observations.

An incomplete list of candidate solutions proposed so far
includes
neutrino oscillations possibly enhanced by Mikheev-Smirnov-
Wolfenstein mechanism \cite{MSW,wolf}, neutrino decay
\cite{decay}, neutrino spin-precession \cite{OVV} in the
solar magnetic field and resonant spin-flavour conversion
scenario \cite{RSF}. Recently, a more sophisticated version
of the last scenario has been produced which employs more
detailed assumptions about the small scale structure of the
solar magnetic field in the convective zone \cite {twist}.

In this work we would like to analyze a simpler and
perhaps more natural hypothesis \cite{p} in connection with
the solar
neutrino problem: the possible existence of nonzero electric
charge
of the neutrino. In fact, the neutrino is the only
elementary particle, besides the gauge bosons, whose
electric
charge is normally assumed to be zero. But if the neutrality
of the gauge bosons is deeply rooted in the principle of
gauge invariance, there are no compelling reasons whatsoever
for the neutrino to have zero charge.

Of course, the neutrino is assumed to be exactly neutral
within the Standard Model. However, the
recently\footnote{Early attempts to study the possibility
and consequences of charge dequantization date back to
Einstein \cite{einstein} who noted that a tiny difference
between the electron and proton charges could account for
the Earth's and Sun's magnetic fields. Later on, Lyttleton
and Bondi \cite{bondi} proposed the idea that such a tiny
difference could cause the expansion of the Universe due to
electrostatic repulsion. It took experimentalists several
decades to rule out these hypotheses (and some of their
modifications).} developed
approach to the problem of the
electric charge quantization has led to the realization of
the fact that in a fairly large class of gauge models,
including the  Minimal Standard Model, the electric charge
can be
dequantized \cite{Mel} (see also \cite{I}). This
means that
the electric charges of elementary particles can take
different values from those conventionally assumed:
 $Q_{\nu}=0$, $Q_{l}=-e, Q_{u,c,b}=2e/3$ and $Q_{d,s,b}=-
e/3$ (e being the modulus of the electronic charge).
In
particular, the neutrino can acquire nonzero electric
charge. (Another interesting aspect of the theories with
dequantized
electric charges is that one might speculate about the
possibility of time dependence of the electric charges
within such theories \cite{we}.)

To be more specific, consider the Standard Model with the
electric charge dequantized through
the formula \cite{Mel}

\begin{equation}
Q=Q_{standard}+\epsilon (L_{e}-L_{\mu}) .\label{a}
\end{equation}
Within this model, the electron and muon neutrinos acquire
electric charges of the same absolute value but of opposite
signs, while the tau neutrino remains neutral:

\begin{equation}
Q(\nu_{e})=-Q(\nu_{\mu})=\epsilon e; \; Q(\nu_{\tau})=0.
\label{b}
\end{equation}
Another possibility is to dequantize charge by a similar
formula \cite{Mel}
\begin{equation}
Q=Q_{standard}+\epsilon (L_{e}-L_{\tau}). \label{c}
\end{equation}
In this case, $\nu_e$ and $\nu_{\tau}$ are (oppositely)
charged while $\nu_{\mu}$ is neutral.
(We would like to emphasize that the above examples are by
no means exhaust all possible ways of giving the neutrino
nonzero charge.)

In other words, in Refs. \cite{Mel} it has been shown that
the Standard Model contains an additional
free parameter $\epsilon$ which must be determined
experimentally along with the other more familiar parameters
such as Higgs mass or Yukawa couplings. Of course if it were
found that $\epsilon$ is nonzero, it should be very small
anyway (see next Section) and that would create one more
hierarchy
problem. Yet taking into account the existence of a few such
problems already, the appearance of a new one does not seem
strong enough argument to disregard the possibility of
nonzero $\epsilon$.

Furthermore, one might argue that nonzero neutrino charge
does not follow from any theoretical principle, whether
established or hypothetical (with the exception of the well-
known rule "all which is not forbidden is allowed"). But
now, based on the works \cite{Mel}, we know that the zero
neutrino charge does not follow from anywhere, too!

Another possible objection against particles with small
fractional charge is that it is difficult to embed them into
grand unified theories \cite{OVZ}. Yet theories with
paraphoton provide a viable alternative \cite{holdom}.

So, at present the cases of zero/nonzero neutrino charges
must be considered as two working hypotheses on the equal
footing, only experiment being capable to provide the
ultimate answer. The situation with neutrino charge is very
similar to the situation with neutrino mass: while zero mass
is the prediction of the minimal standard model, most
physicists agree that the question of zero/nonzero neutrino
mass has much more to do with experimenting than with model-
building. While it would not be easy to detect the neutrino
charge, the consequences of such discovery should certainly
be dramatic, ranging from the prospects of detecting relic
neutrino through its electromagnetic interaction to possible
better ways of managing neutrino beams, creation of neutrino
optics etc.

Finally, let us note that in the present work we are not
concerned if the neutrino mass is zero or not. Certainly,
there exist well-known difficulties associated with
charged massless particles \cite{massless}. However, one can
take a pragmatic point of view \cite{brf} and keep
developing a theory until one runs into any inconsistency.
No such inconsistency seems to show up in our treatment.
An alternative point of view is to give the neutrino a Dirac
mass by introducing additional Higgs multiplets.

Note also that even if the neutrino is massless in vacuum,
it cannot be considered massless inside plasma. This is
because the vacuum dispersion relation $E=| {\bf p}|$ gets
changed by the weak interaction of neutrino with plasma
\cite{wolf}. In other words, there arises a refraction index
for the neutrino propagating through plasma. Thus, the
situation with infrared divergencies might be better for a
neutrino in plasma than in vacuum.

To conclude, it seems that assuming nonzero neutrino charge
is certainly not more heresy than assuming nonzero neutrino
magnetic moment, or mass and mixing angles.

In addition, there exist quite an independent motivation to
study the behaviour of a charged neutrino inside the Sun.
The point is that the neutrino electromagnetic properties
get modified by plasma effects and under certain conditions
these modifications result in {\em an induced electric
charge of the neutrino\/} \cite{induced}. We stress that it
happens in the Minimal Standard Model where the neutrino has
zero {\em intrinsic\/} electric charge.

The purpose of  this paper  is to draw attention to the fact
that the possible existence of a very small  electric charge
of the neutrino may be a clue to the solar neutrino
problem\footnote{The possible role of the charged neutrino
interaction with the terrestrial electric field, in
connection with the solar neutrino problem, was discussed
previously by G.C.Joshi and R.R.Volkas (unpublished).}. The
idea is that the charged  neutrinos  are
deflected by the solar  magnetic field  while passing
inside the  Sun. Thus the resulting neutrino flux is made
anisotropic which leads to the solar neutrino deficit
registered on the Earth. (Within our scenario, this deficit
is not real but only apparent in the sense that {\em the
total \/} $4 \pi$ solar neutrino flux is {\em not changed\/}
as compared with the standard solar models.)

Section 2 summarizes the existing bounds on the electron
neutrino charge. In Section 3 we give an order-of-magnitude
estimate of the effect. Section 4 is devoted to the more
detailed calculation of the solar neutrino deficit in the
present scenario. In Section 5 some experimental
implications are discussed and one of them - the prediction
of a second neutrino flux - is briefly outlined in Section
6. Our main results are summarized in Section 7.

\section{phenomenological limits on the electric charge of
the electron neutrino}
For a recent detailed analysis of various bounds on the
neutrino charges see \cite{bv}.

The strongest model-independent\footnote{By model-
independent we mean the constraints that do not rely on
additional assumptions such as charge conservation or the
equality $Q(\nu_{e})=Q(\bar{\nu_{e}})$.} constraints on the
electron
neutrino charge, $Q(\nu_{e})= \epsilon e$, come from three
sources.

1) Analysis of the data on $\nu_{e} e$ elastic scattering
\cite{brf}:

\begin{equation}
\epsilon \alt 3 \times 10^{-10}.
\end{equation}

2) Study of the electromagnetic neutrino production in
the core of the Sun and its influence on the solar
energetics \cite{brf}:

\begin{equation}
\epsilon \alt 10^{-13}. \label{sun}
\end{equation}
This estimate follows from the fact that if the neutrino is
charged then the neutrino-antineutrino pairs will be
produced in the decays of plasmons (which can be viewed as
massive photons) in the core of the Sun. The neutrino will
then escape freely from the Sun thus taking away a lot of
energy which would be inadmissible in terms of the solar
energy balance.

3) Analysis of the data on SN1987A supernova explosion
\cite{bc}:

\begin{equation}
\epsilon \alt 10^{-(15 \div 17)}. \label{SN}
\end{equation}

These upper limits on the electron neutrino charge
$\epsilon$ were based on the experimental detection
\cite{hirata1987} of the neutrino signal from the supernova
1987A explosion. It was claimed that if $\epsilon$ were
larger than (\ref{SN}) then the intergalactic and galactic
magnetic fields would lenghthen the neutrino paths, and
neutrinos of different energy could not arrive on the Earth
within a few seconds of each other, even if emitted
simultaneously by the supernova.

Yet it is generally believed (see, e.g.\cite{partdata}) that
the constraint based on SN1987A arguments, although
stronger, is less reliable than the previous ones because it
involves the
details of the galactic magnetic field which are not very
well known.

There exist even more severe, but less direct constraints.
They  are based on the experimental data on the neutrality
of
atoms and neutrality of the neutron. These data give limits
on the sum of the proton and electron charges \cite{ep}:
$Q(p)+Q(e)= (0.8 \pm 0.8) \times 10^{-21}e$ and the neutron
charge \cite{n}: $Q(n)= (-0.4 \pm 1.1) \times 10^{-21}e$.
Then, assuming charge conservation in the neutron beta-decay
$n \rightarrow p+e^{-}+ \bar{\nu_{e}}$  we can obtain the
bound on the electron {\em anti}neutrino charge: $ Q(
\bar{\nu_{e}}) < 3 \times 10^{-21}e$. Finally, assuming
validity of CPT symmetry\footnote{Besides charge
conservation and CPT,  a number of usually unspoken but very
important assumptions underlying the constraint (\ref{-21})
are made. For instance, one has to assume that the electric
charges of {\em free\/} electrons and protons are exactly
the same as those of {\em atom-bound\/} electrons and
protons. Another fundamental assumption, as noted in Ref.
\cite{brf}, is that the electric charge, as measured by
interaction with an electromagnetic field, coincides with
the electric charge assigned by the charge-conservation law
(see also a discussion of that point in \cite{fg}).
According to Ref. \cite{brf}, it is possible to construct
models in which it is not the case.  Under ordinary
circumstances there is no doubt in the correctness of the
above axioms, but when it comes to such fantastic accuracies
as $10^{-21}$, it does not seem unreasonable to question
those axioms, too.} with respect to $\nu_{e}$ and $
\bar{\nu_{e}}$ charges, one can claim that

\begin{equation}
 Q(\nu_{e}) < 3 \times 10^{-21}e. \label{-21}
\end{equation}
In particular, this constraint is valid within the model
with charge dequantization through Eq.~(\ref{a}) and
Eq.~(\ref{c}). Yet the requirements of the electric charge
conservation and CPT symmetry, although very general and
perfectly valid up to now, are themselves a subject of
experimental testing at present\footnote{Note that it is
possible to constrain the $\nu_{e}$ charge assuming only
charge conservation in the decay $n \rightarrow
p+e^++\nu_{e}$, but not the equality
$Q(\nu_{e})=Q(\bar{\nu_{e}})$. Naturally, this constraint
turns out to be much weaker than (\ref{-21}), namely:
$Q(\nu_{e}) < 4 \times 10^{-8}e$ \cite{hughes}.}.
Furthermore, there exist
several models in which charged neutrino arises as a natural
consequence of the electric charge violation \cite{Qnoncons}
(in those specific models the constraint (\ref{-21}) is
still not invalidated).

In this work we are not concerned
with the problem of constructing a viable model which would
allow one to avoid the constraint (\ref{-21}). Instead,
in the main part of the paper we do not use any model-
dependent arguments and we treat the neutrino charge
$\epsilon$ as a free parameter; also we consider some
interesting consequences of the choice $\epsilon \simeq
10^{-13}$.
Of course, should any of the above constraints change in the
future, our results would be easily reformulated by simply
rescaling the magnitude of $\epsilon$.

A massive Dirac $\nu_{e}$ with an electric charge of $10^{-
13}e$ and a mass of $< 7 \div 9$ {\rm ~eV}  \cite{numass}
would have a Dirac magnetic moment $\mu > 5 \times 10^{-9}$
$ \mu_B$, where $\mu_B$ is the Bohr magneton. Therefore one
might suspect a contradiction with the existing upper bounds
on the neutrino magnetic moment obtained recently by a
number of authors (see, e.g., \cite{raffelt} and references
therein): $\mu (\nu_e) < 10^{-(10 \div 12)}$ $\mu_B$.
However, these limits apply only to the {\em anomalous\/}
magnetic moment, but not to the Dirac magnetic moment. In
our case, the neutrino anomalous magnetic moment due to the
electromagnetic radiative correction is equal to $\epsilon
^2 \alpha /2 \pi$ and therefore negligible. Still, some of
the constraints on $\mu (\nu_e)$ can be translated into
limits on $Q(\nu_{e})$. For example, the analyses of plasmon
decay into neutrino-antineutrino pair due to $\mu (\nu_e)$
in various astrophysical contexts can be carried out also
for the case of a charged neutrino. Yet, the resulting
limits are not significantly different from the solar
constraint (\ref{sun}).

Note also that recently there has been considerable interest
in discussing the possible existence of new particles
carrying very small electric charge  ("milli-charged
particles") \cite{I}. These works contain detailed
discussion of many phenomenological constraints on such
particles obtained from a variety of sources (including
astrophysics, cosmology, geophysics and macroscopic
electrodynamics). Many of those constraints apply to the
case of electron neutrino, too; we shall not repeat that
material here.

\section{Simple estimate }
Since the neutrino charge, if any, must be so tiny, $ \alt
10^{-13}e$, at first sight it seems unlikely that it would
play any role at all. To make a rough check whether it is
true or not, let us start by a dimensional order-of-
magnitude estimate of the effect of the solar magnetic
field\footnote{Here, by "the solar magnetic field" we mean
the
large scale toroidal magnetic field in the convective zone.
For the moment, we ignore the magnetic fields in the core
and the radiative zone - not because they are irrelevant but
because we do not know much about them.} on the the charge
neutrino motion inside the Sun. As a dimensionless
characteristic of the effect it is natural to choose the
neutrino deflection angle.  The deflection angle $\gamma$
may be written as the product of the deflection rate, $d
\gamma/dt$, and the time of flight through the convective
zone, $\tau$:

\begin{equation}
\gamma \sim {d \gamma \over dt} \times \tau, \; \tau \simeq
{D
\over c}.
\end{equation}
where $D \simeq 2 \times 10^{10} cm$ is the thickness of the
convective zone, $c$ is the speed of light. Now, in a crude
approximation, the value of the deflection rate $d
\gamma/dt$ will depend only on the neutrino electric charge,
$\epsilon e$, the neutrino energy $E \simeq 1 {\rm ~MeV} $
and some characteristic value of the solar magnetic field,
$H_{c}$. The only combination of dimension $time^{-1}$ which
can be made out of these quantities is the Larmour
frequency,

\begin{equation}
\omega_L = {\epsilon e c H_{c} \over E},
\end{equation}
(We use the Gauss system of units throughout the paper,
i.e., $e=4.8 \times 10^{-10}$ esu, 1 {\rm ~MeV} $=1.6 \times
10^{-6}$ erg.)
Putting all together and adopting $\epsilon \simeq 10^{-13}$
for the neutrino charge and $H_c \simeq 10^{4} \div 10^{5}
G$ for the solar magnetic field\footnote{This choice is
discussed in more detail below, between Eqs.~(\ref{13}) and
(\ref{131}).} we finally obtain for the
deflection angle

\begin{equation}
\gamma \sim {\epsilon e H_{c} D \over E} \simeq  0.01 \div
0.1. \end{equation}
Thus we see, that contrary to a naive expectation, the
possible tiny charge of the neutrino may indeed play some
role in the neutrino propagation through the solar magnetic
field. Yet from our estimate it is hard to see the relation
between the neutrino deflection angle and the neutrino
deficit observed on the Earth. Also, the above consideration
suggests that the seasonal variations of the solar neutrino
flux must be large (because of two facts: first, the Earth's
orbit is inclined with respect to the solar equator; second,
the solar magnetic field is small at the equator and it
grows away from the equator). It will turn out that this
suggestion is {\em wrong\/} and that seasonal variations in
our scenario would not be significant. So it is worthwhile
to study
the problem more carefully.

\section{Calculation of the neutrino flux on the Earth}

We have seen, by an order-of-magnitude estimate, that the
suggested value of the neutrino charge , $\epsilon \simeq
10^{-13}$, may be relevant to the solar neutrino problem.
Now we
can start a more exact calculation of the reduction of the
solar neutrino flux at Earth due to the bending of neutrino
trajectories by the solar magnetic field. The key quantity
we need to find is the angular distribution of the neutrino
flux coming out from the surface of the Sun assuming that
the initial flux
coming out from the central region of the Sun is isotropic.

Therefore, the procedure is three step.
First, we start with the isotropic angular distribution.
Second, we find out how the neutrino trajectories are bent
by the solar magnetic field. Mathematically, this bending of
trajectories can be viewed as a change of variables in the
distribution function. The last step, then, is to make
explicitely such change of angular variables that would
lead us to the final angular distribution. From this
anisotropic distribution we shall easily obtain our key
result, Eq.~(\ref{12}) : the reduction of the neutrino flux
predicted by the present hypothesis (and then compare it
with the reduction of the flux observed experimentally).

For definiteness and simplicity of the calculation, we make
the following assumptions.
First, we assume that all the neutrinos are emitted from a
point
source located in the centre of the Sun. In other words we
neglect the radius of the neutrino-emitting zone, $R_{\nu}$,
as compared
to the solar radius $R_{\odot}$ (their ratio is
$R_{\nu}/R_{\odot} \simeq 0.1$).
Next, we suppose that the neutrino deflection angle is small
(the precise meaning of that assumption will be discussed
below).
Furthermore, what we are interested in is not the complete
$4\pi$ angular distribution of the neutrino flux but only
a small part of it within the angular interval swept by the
line of vision Earth-Sun.  This is a narrow interval close
to the solar equatorial plane: $-7^o < \theta < 7^o$.  This
fact will be used throughout the calculation to simplify it.

Finally, a few remarks about the solar magnetic field. The
structure of the solar magnetic field is very complicated
and is probably one of the worst known aspects of the Sun.
Yet, some important information about it can be obtained
from the observations of  solar
activity such as sunspots which strongly depends on the
behaviour of the large- scale magnetic field inside the Sun.
For instance, observing Zeeman effect for the light emitted
from the sunspots one can measure the strength of the
magnetic field near the surface of the Sun which can then be
translated into estimates of the asimuthal magnetic field in
the convective zone of the Sun.
In what follows we will rely only on relatively well
understood features of the solar magnetic fields which
include (see, e.g. \cite{parker}):

1)the existence of the large-scale toroidal magnetic field
in the convective zone of the Sun (toroidal means the
magnetic lines of force are closed circles parallel to the
Sun's equatorial plane);

2)the directions of the toroidal fields in the nothern and
southern hemispheres of the Sun are opposite. Besides, these
directions reverse themselves every 11 years;

3)the strength of the toroidal magnetic field is nearly zero
in the equatorial plane and it grows with the separation
from the equatorial plane within the narrow transition zone,
between about $+10^o$ and $-10^o$ latitudes.
Therefore, in this tranzition zone there exists a gradient
of the toroidal magnetic field in the direction parallel to
the rotation axis of the Sun. Unfortunately, the magnitude
of the gradient is not very well known and we will consider
it as a parameter varying in a certain range (see below,
between Eqs.~(\ref{13}) and (\ref{131}).
It is this gradient rather than magnetic field itself that
plays the key role in reducing the neutrino flux on the
Earth within the present scenario.

According to our plan, we start with the isotropic neutrino
flux from the centre of the Sun described by the angular
distribution
%1
\begin{equation}
{dN \over d\Omega} = N_0 = const,
\end{equation}
or equivalently,
%2
\begin{equation}
F_{0}( \theta, \phi) \equiv  {dN \over d \theta d \phi} =
\sin{\theta},
\end{equation}
where $\theta$ and $ \phi $ are the usual angles in the
spherical coordinate frame whose origin is placed in the
centre of the Sun, y-axis points to the Earth and z-axis is
the rotation axis of the Sun.
After passing the convective zone filled with magnetic field
the neutrino is deflected from the direction characterized
by the pair of spherical angles $\theta$ and $ \phi $ to the
new direction characterized by the spherical angles $\alpha$
and $\beta$. Using the equation of motion, we can
explicitely write down the angles $\alpha$ and $\beta$ as
functions of $\theta$ and $ \phi$. Conversely, we can
express $\alpha$ and $\beta$ as functions of $\theta$ and $
\phi$:

\begin{equation}
\theta=T(\alpha,\beta), \phi=P(\alpha,\beta).
\end{equation}
Now, all we need to get the new distribution function from
the old one is to change variables from $\theta$ and $\phi$
to $\alpha$ and $\beta$. Then, the new distribution function
which we denote by $F_1(\alpha,\beta)$ will look like

\begin{equation}
F_1(\alpha,\beta)=F_0(T(\alpha,\beta),P(\alpha,\beta))
J(\alpha,\beta), \label{0}
\end{equation}
where $J(\alpha,\beta)$ is the Jacobian
%3
\begin{equation}
J( \alpha, \beta)= \left|\begin{array}{cc} { \partial T
\over \partial \alpha} & {\partial T \over \partial \beta}\\
{\partial P \over \partial \alpha}&{\partial P \over
\partial \beta} \end{array}\right|. \label{8}
\end{equation}
Finally, within our model, the neutrino flux observed on the
Earth ($\Phi_1$) and the flux predicted by the standard
solar model ($\Phi_0$) are related simply by

\begin{equation}
{\Phi_1 \over \Phi_0}={F_1(\pi/2,\pi/2) \over
F_0(\pi/2,\pi/2)}. \label{11}
\end{equation}

Now, we proceed to the calculation of the angular
distribution function $F_1$. To do that,  we need to find
the deflection angles, that is the functions
$T(\alpha,\beta)$ and $P(\alpha,\beta)$. For this purpose,
let us consider the equations of motion for a charged
neutrino in the magnetic field

%4
\begin{equation}
{d {\bf p} \over dt} = {\epsilon e \over c} [{\bf v} \times
{\bf H}],\; {dE \over dt}=0,
\end{equation}
where $ \bf p$ and $\bf v$ are the neutrino's
momentum and velocity, respectively, $\epsilon e$ is the
neutrino electric charge.
Since the electron neutrino mass is experimentally
known to be less than $7 \div 9 $ {\rm ~eV} \cite{numass},
to a
very high accuracy the neutrino velocity equals the velocity
of light, $v \approx c$. As for the neutrino energy, $E$,
its magnitude depends on the reaction in which the neutrino
is born. On the other hand, it must be greater than the
experimental threshhold which is different for different
experiments, ranging from  about 0.2{\rm ~MeV} to 7 {\rm
{}~MeV}.

To simplify the equation of motion, recall that we are only
interested in what happens in a close vicinity of the
direction Sun-Earth, that is near the direction
$\theta=\pi/2, \phi=\pi/2$. Therefore we can neglect the
curvature of the magnetic lines of force and of the
convective zone containing these lines of force and
approximate the convective zone by a flat slab of the same
thickness, $D \simeq 2 \times 10^{10}cm$, containing the
{\em
straight \/}  lines of force. This slab is placed
perpendicular to the axis Earth-Sun. The distance between
the centre of the Sun and the middle of the slab is taken to
be $d \simeq 6 \times 10^{10}$cm - which is the distance
between the
centre of the Sun and the middle of the convective zone.
Furthermore,  we assume that the magnetic field is zero
outside this slab whereas inside the slab the only nonzero
component is $H_x$ (i.e., we ignore the poloidal magnetic
field altogether). In the absence of a detailed model for z-
dependence of the magnetic field, we adopt the simplest
form, i.e. a linear variation with z:
%5
\begin{equation}
H_y=H_z=0, \; H_x= \langle {\partial H_x \over \partial z}
\rangle z.
\end{equation}
where  $ \langle {\partial H_x \over \partial z} \rangle $
is the average vertical gradient of the large-scale magnetic
field.

Since the deflection angles are assumed to be small, we can
neglect the variation of magnetic field over the trajectory
of any single neutrino flying at some
$\theta$. In other words, we assume that all the neutrinos
flying at the {\em same \/} angle $\theta$ will feel {\em
the same\/} strength of the magnetic field, while the
neutrinos flying at {\em different\/} angles $\theta$ will
feel {\em different\/} strengths of the magnetic field. To
paraphrase it once more, we assume that a neutrino flying at
an angle $\theta$ upon entering the convective zone will
move in the {\em uniform\/} magnetic field all the way
through until it leaves the Sun; the strength of that
uniform magnetic field is taken to be the same as the
strength of the magnetic field at the "entrance point":
%6
\begin{equation}
H_y=H_z=0, \; H_x= \langle {\partial H_x \over \partial z}
\rangle d \cot
{\theta} \approx \langle {\partial H_x \over \partial z}
\rangle d ( { \pi \over 2} - \theta).
\end{equation}

Now, the motion of a charged particle in a uniform magnetic
field is well-known: along the magnetic field the motion is
free, while in the perpendicular plane the trajectories,
both in coordinate and momentum spaces, are circles orbited
with Larmour frequency. Thus, if a neutrino enters the
convection zone with the speed components $v_{0x}, v_{0y},
v_{0z}$, it will leave the Sun with the speed components

\begin{equation}
v_ x =v_{0 x }, \label{1}
\end{equation}
\begin{equation}
v_ y =v_{0 y }\cos{\omega \tau}+v_{0 z }v_{0 z }\sin{\omega
\tau} \approx v_{0 y } + v_{0 z } \delta, \label{2}
\end{equation}
\begin{equation}
v_ z =-v_{0 y }\sin{\omega \tau}+v_{0 z }\cos{\omega \tau}
\approx -v_{0 y } \delta + v_{0 z }. \label{3}
\end{equation}
Here, $ \tau=D/c $ is the time of flight through the
convection zone\footnote{Note that the actual times of
flight, besides being different for different neutrinos, are
in fact longer than $D/c$ because the neutrino paths are not
straight, but this difference is negligible (and even if it
were not, assuming $ \tau=D/c $ for all neutrinos would only
decrease the effect rather than increase it).},
$\delta=\omega \tau$ is the angle by which the magnetic
field rotates the neutrino speed vector projected to yz-
plane; $\omega$ is the Larmour frequency:
%7
\begin{equation}
\omega= \omega_0 \cot {\alpha} \approx ( {\pi \over 2} -
\alpha), \; \omega_0 = { \epsilon e c
\over E}  \langle {\partial H_x \over \partial z} \rangle d.
 \label{4}
\end{equation}

{}From these equations we see that the condition of validity
of our small-angle approximation can be formulated as
$\sin{\delta} \approx \delta$.

Having obtained the solution of the equations of motion, we
are now able to find the relation between the initial angles
$\theta,\phi$ at which a neutrino enters the convective zone
and the final angles $\alpha, \beta$ at which the neutrino
leaves the Sun. For this purpose, let us express the initial
speed components $v_{0x}, v_{0y}, v_{0z}$ through the
initial angles $\theta,\phi$, and the final speed components
$v_{x}, v_{y}, v_{z}$ through the final angles $\alpha,
\beta$:
\begin{equation}
v_{0x}=c \sin{\theta}\cos{\phi}, \; v_{0y}=c
\sin{\theta}\sin{\phi}, \; v_{0z}=c \cos{\theta}. \label{5}
\end{equation}
\begin{equation}
v_{x}=c \sin{\alpha}\cos{\beta}, \; v_{y}=c
\sin{\alpha}\sin{\beta}, \; v_{z}=c \cos{\alpha}. \label{6}
\end{equation}
Assuming as usual that the deflection angles are small,
we can write

\begin{equation}
\theta=\alpha+\Delta\alpha, \; \phi=\beta+\Delta\beta,
\label{7}
\end{equation}
where $\Delta\alpha$, $\Delta\beta$ are small.
Now, solving the system of equations,
Eqs.~(\ref{1}) - (\ref{7}) with respect to $\Delta\alpha,
\Delta\beta$, we find

\begin{equation}
\theta=\alpha-\delta_0\sin{\beta}\cot{\alpha} \equiv
T(\alpha,\beta) \approx \alpha - \delta_0 ({\pi \over 2}-
\alpha),\label{9}
\end{equation}
\begin{equation}
\phi=\beta- \delta_0\cot^{2}{\alpha}\cos{\beta} \equiv
P(\alpha,\beta) \approx \beta, \label{10}
\end{equation}
where $\delta_0=\omega_0 \tau$.
Having expressed initial angles in terms of final
angles, that is, having obtained the functions
$T(\alpha,\beta)$, $P(\alpha,\beta)$ we can now calculate
the
Jacobian in Eq.~(\ref{8}) and then insert the result,
together with  Eqs.~(\ref{9}), (\ref{10})  into our basic
formula,
Eq.~(\ref{0}). Lastly, we recall Eq.~(\ref{11}) giving the
neutrino flux observed on the Earth within our model.

Thus, finally, the neutrino flux observed on the Earth
within our model is given by

\begin{equation}
{\Phi_1 \over \Phi_0}= 1+ \delta_0= 1+ {\epsilon e \langle
{\partial H_x \over \partial z} \rangle d D \over E} .
\label{12}
\end{equation}

An important thing about this result is that it is {\em
the field gradient \/} rather than {\em magnetic field
itself \/} that causes the effect. Indeed, using our method,
one can show that if the toroidal  magnetic field in
the convective zone were {\em uniform\/}, it would not lead
to any
substantial anisotropy of the neutrino flux at all.

To compare  Eq.~(\ref{12})  with the experimental data, we
need to know the value of the mean gradient (along the solar
rotation axis) of the large-scale toroidal magnetic field in
the convective zone ,  $ \langle {\partial H_x \over
\partial z} \rangle $. Unfortunately, the magnitude of this
gradient is not very well known, so let us try to reverse
our
problem and ask: what value of the gradient will be needed
for our mechanism to explain the solar neutrino deficit?
Assuming the neutrino charge to be $\epsilon = 10^{-13}$,
its energy $E=0.8$ {\rm ~MeV}
and requiring the flux deficit to be $\delta_0=-0.5$, from
Eq.~(\ref{12}) we find that the gradient must be
\begin{equation}
\langle {\partial H_x \over \partial z} \rangle \approx -1.1
\times 10^{-5} \; G/cm. \label{13}
\end{equation}
Is it a reasonable figure or not? A crude estimate of the
gradient can be obtained by dividing $H$ by $h$ where $H$ is
the maximum value of the magnetic field reached at the
latitudes of about $ \pm 10^o$ \cite{parker} and $h$ is the
distance from that latitude to the solar equatorial plane,
$h=d \sin{10^o} \approx 10^{10}$cm.

As for the possible
value of $H$, it is a subject of a debated controversy.  On
the one hand, it is claimed \cite{smallH} that values of $H$
greater than $10^{4} G$ are ruled out by the non-linear
growth-limiting effects; on the other hand, there are
arguments based on the helioseismology data that it can
reach as large values as a few million G \cite{largeH}.
Anyway, magnetic fields up to $10^{4}$ G (or even $10^{5}$ G
\cite{akh}) are widely used by many authors trying to
explain the solar neutrino puzzle. So, we leave it to the
reader to make his/her own judgement on that point. Note
also, that it is the magnetic field close to the surface of
the Sun which reaches its maximum at $10^o$ latitude, and
this latitude may be higher (or lower) for magnetic fields
located at larger depths. That brings in an additional
uncertainty to the estimate of the gradient. If we do admit
that the magnetic field in the convective zone may vary in
the range $H= 10^{3} \div 10^{6}G $ than the value of the
gradient may vary in the range

\begin{equation}
\langle {\partial H_x \over \partial z} \rangle  \simeq {H
\over h} \approx (10^{-7} \div 10^{-4}) \; G/cm .\label{131}
\end{equation}
Hence we see that the value of gradient needed  to explain
the neutrino deficit, Eq.~(\ref{13}) {\em may indeed exist
in the convective zone of the Sun\/}.

A few remarks are now in order.

1.We have just considered the case of $\alpha= \pi/2$
corresponding to the case when the Earth traverses the plane
of the solar equator. But that
happens only twice a year: in June and December - the days
of
summer and winter solstice. During the rest of time the
direction
Sun-Earth is at an angle to the solar equatorial plane.
Fortunately,
this angle is rather small: $|\pi/2 - \alpha| \alt 7^o $, so
that our small-
angle approximation still applies.
(The boundaries are reached when the Sun - Earth axis lies
in the northern solar hemisphere at an angle of $7^o$ in
September,
autumn equinox, and through the southern solar hemisphere,
at the
same angle, in March, spring equinox.) If we keep $\alpha$-
dependence in our calculation then we obtain that the
deficit of the neutrino flux varies according to

\begin{equation}
{\Phi_1 \over \Phi_0} = 1+{\delta_0 \over \sin^{2}{\alpha}
}. \label{14}
\end{equation}
Thus for extreme values of $\alpha$ we have\footnote{Note
that Eq.~(\ref{14}) cannot be extrapolated to large angles
so it does not mean that the neutrino flux is zero at
$\alpha = \pi$ and $\alpha=0$.}:

\begin{equation}
{\Phi_1 \over \Phi_0}=1+{\delta_0 \over \sin^{2}{97^o}}=1+
{\delta_0 \over \sin^{2}{83^o}}
\approx 1+1.01 \times \delta_0.
\end{equation}
Recall that this result is valid within our system of
approximations: we neglected the curvature of the solar
convective zone and substituted the true value of the
magnetic field gradient by its average value. But even if
these assumptions are dropped, there are hardly any reasons
to expect that this result would change dramatically.

So we can conclude that the seasonal variations of the
neutrino flux
predicted by the present model are rather small which makes
them very hard to observe in the experiment.

2.All the above derivation was of course completely
classical. Indeed, one can reasonably expect that quantum
corrections (e.g., due to neutrino diffraction) must be
vanishingly small.

3. One might wonder about possible influence of other
magnetic fields encountered by the neutrinos on their long
way from the Sun to the Earth. Specifically, there are the
solar magnetic field outside the Sun, the interplanetary
magnetic field and, finally, the terrestrial magnetic field.

First of all, the terrestrial magnetic field is completely
transparent for solar neutrinos because the penetration
ability of a particle of charge $q$ and momentum $p$ is
controlled by the factor $pc/q$ which is huge for a neutrino
with the charge $ 10^{-13} $ - it is by many orders of
magnitude greater than this factor for the penetrating
particles of cosmic rays. The neutrino deflection angle due
to Earth's magnetism is also negligibly small since both the
terrestrial magnetic field and the time of neutrino flight
through the terrestrial magnetosphere are much less than
those for the convective zone of the Sun.
Similarly, the influence of the outer magnetic field of the
Sun and the interplanetary magnetic field can be neglected:
although the time of flight outside the Sun is about
$10^{3} $ longer than inside, the magnetic field outside is
$ 10^{-4} \div 10^{-5} $ G i.e., many orders of magnitude
weaker than
inside, so that the product of these two factors outside the
Sun is much smaller than inside. In addition to that, let us
stress once
again that we need large {\em field gradients\/} rather than
large {\em magnetic fields\/} which are unlikely to emerge
in the interplanetary space.

4. Also, the role of various {\em electric\/} fields, both
inside and outside the Sun, seems to be negligible for the
present problem. Inside the Sun, as is well known for any
plasma, any electric field dissipates very quickly due to
high conductivity of the plasma\footnote{In fact, there may
exist exceptions to this rule. Under certain circumstances
(for example, in the conditions of solar flares) the
electrical resistivity may be greatly increased so that a
local electric field may arise \cite{parker}. This point is
probably worth further studying.}. Outside the Sun, the
electric field could exist, for instance, due to nonzero
total electric charge of the Sun. Although there is not much
information available about the strength of such electric
field, we can be quite confident that it cannot be strong
enough to capture the neutrinos with the charge $ 10^{-13}e$
and the energy $E \sim 1 {\rm ~MeV}$ before they reach
the Earth (or even decrease their energy to any extent). If
that was possible, then the solar wind could never reach
close to the Earth, too. The terrestrial electric field,
about 1 V/cm in strength, is also helpless to stop the
neutrinos.

5. One might wonder if the nonzero neutrino charge would
affect the solar neutrino detection at Kamiokande. The point
is that there arises an electromagnetic contribution to the
cross section of the neutrino-electron scattering which
might modify the Kamiokande results. However,  if
$Q_{\nu}=10^{-13}e$ then the cross-section of the
electromagnetic $\nu_e e$ scattering is about 10 orders of
magnitude smaller than the cross-section of weak $\nu_e e$
scattering and therefore can be completely neglected.

6. At this stage, we have not considered the effect of solar
matter on our results.

7. Now we come to the discussion of the most serious flaw of
the suggested scenario in its present form. The point is
that up to now we tacitly assumed that the direction of the
magnetic field is such that the value of the deficit,
$(-\delta_0)$, is positive. But recall that the large-scale
magnetic field in the convective zone {\em reverses\/} every
11 years which means that the gradient, too, changes its
sign every 11 years. That means that taken as it is,
Eq.~(\ref{12}) would predict that each 11 year period of
neutrino flux {\em deficiency\/} must be followed by 11 year
period of neutrino flux {\em excess of the same magnitude\/}
so that the flux averaged over the 22 year cycle would be
the same as predicted by the Standard Solar Model.

One can think of several possible ways out of that
difficulty.

The most natural one is to recall that our previous
calculation was based on the approximation of small
deflection angles, or, more precisely, the smallness of the
parameter $\delta_0$. We can expect that this small-angle
approximation is valid as long as $ \sin{\delta_0} \approx
\delta_0$ or $|\delta_0| \alt 0.7$. But if the gradient of
the
magnetic field is greater than $ 2 \times 10^{-5}$ G/cm than
our
approximation does not work anymore and a more exact
calculation is needed.

Naively, one might expect that the neutrino deficiency must
alternate with the neutrino excess at 11 year intervals
independently of the magnitude of the gradient: just note
that  when the magnetic field configuration is defocusing,
one would expect the neutrino deficiency and when it is
focusing, the neutrino excess. Each reversal of the magnetic
field
means a switch between focusing and defocusing modes so that
any 11 year "deficiency" cycle would be followed by the 11
year
"excess" cycle, however great the gradient of the magnetic
field is. Nevertheless, there are arguments based on simple
geometrical optics considerations which show that it is not
the case
and if the gradient is large enough then the neutrino
deficiency can occur both for the defocusing {\em and the
focusing\/} configuration!

Another option is to try to relax the solar upper bound on
the possible electric charge of the electron neutrino
obtained in \cite{brf}, since the neutrino charge and the
magnetic field gradient come always as the product $\epsilon
\times \langle {\partial H_x \over \partial z} \rangle $
rather than separately. Note that if we used only the most
reliable bound (i.e., direct experimental bound extracted
from the elastic $\nu_{e} e$ scattering data) on the
electron neutrino charge, $ \epsilon \alt 3 \times 10^{-10}$
\cite{brf}, the required value of the gradient would be less
than the above estimate by the factor of 300.

Let us also mention briefly that at present we cannot rule
out the possible existence of a primordial magnetic field of
as much as $10^{6}$ G inside the core of the Sun
\cite{parker}. Within the present context, it would very
interesting if any evidence could be obtained concerning the
existence of significant gradients of that field near the
 plane of the solar equator.

Finally, although it is not as much appealing, we should not
discard the possibility that our mechanism is effective only
during alternative 11 cycles or even only during the periods
of active sun within the alternative 11 year cycles while
some other mechanism is responsible for neutrino depletion
during the rest of the time. This possibility will have to
be considered much more seriously if the anticorrelation of
the neutrino deficiency with solar activity is established
firmly by the future experiments.

\section{DISCUSSION}

As we stressed in the preceding sections, our present
knowledge of the structure of the solar magnetic field is
rather limited. Thus, we cannot rule out such values of the
magnetic field that would lead to {\em large\/} deflections
of the
neutrinos travelling through the convective zone of the sun.
The quantitative theory for this case seems more difficult
to construct and we do not attempt it here. Yet, it is
instructive to discuss here some qualitative features of
such theory based on simple physical considerations, keeping
in mind the results of four different solar neutrino
experiments available by now.

Before doing that, let us very briefly summarize the solar
neutrino data.

1) Anticorrelation of the neutrino flux with solar activity
is probably observed in the Homestake data \cite{anti,bapr}.

2)No such anticorrelation is observed in the Kamiokande data
\cite{kamioka}.

3)The higher neutrino flux (i.e., less neutrino deficit) is
observed in Kamiokande experiment than in Homestake
experiment.

4)The higher neutrino flux is observed in SAGE \cite{sage}
and GALLEX \cite{gallex} experiments than in Homestake
experiment.

It is very important for us to note that the experimental
thresholds of neutrino energy are rather different in those
experiments: $E_{Home}=0.816$ {\rm ~MeV}, $E_{Kam} \sim 7.5$
{\rm ~MeV}, and $E_{Gallex}=0.233$ {\rm ~MeV}.

Now, let us discuss qualitatively some effects that are
consequences of our hypothesis. These effects are controlled
by the neutrino electric charge $\epsilon$, gradient of the
solar magnetic field $\langle {\partial H_x \over \partial
z} \rangle$, and the neutrino energy $E$ (for the moment,
we forget about other relevant parameters). However, these
quantities enter not separately but only through the ratio
$\epsilon \langle {\partial H_x \over \partial z} \rangle
/E$. Therefore, changing the magnetic field will be
equivalent to changing the neutrino energy
correspondingly.
Furthermore, it is natural to assume that both neutrino flux
deficit and anticorrelations grow with the increase of the
gradient. Hence we obtain that the anticorrelations have to
be {\em smaller\/} for {\em more energetic \/} neutrinos.
And this is exactly what is needed to qualitatively explain
the difference between Homestake and Kamiokande data (see 1)
and 2) above). Also, by the same reasoning, within our
scenario one can expect less deficit in Kamiokande than in
the Homestake experiment.
Thus we can summarize that it is rather plausible that the
present scenario can, in principle, account for the three
out of four main experimental features, see 1)-3) above.

Now, as for the fourth feature, i.e., results of gallium
experiments, our hypothesis seems to predict {\em greater\/}
deficit than Homestake and thus looks disfavored by gallium
results. However, one must remember that: 1) the difference
between Gallium and Homestake results, from the viewpoint of
our hypothesis, must be {\em less\/} pronounced than the
difference between Homestake  and Kamiokande data. This
follows from the fact that the ratio of the characteristic
neutrino momenta for Homestake-Gallium data are {\em
smaller\/} than for Kamiokande-Homestake data;
2) The errors of Gallium data are still larger than those of
Homestake data.

Now, we would like to draw attention to a curious
coincidence in the solar neutrino data. Kamiokande does not
see anticorrelations during the whole period of its
operation, i.e., 1987-1993  (part of
solar cycle \# 22). And, according to \cite{bapr}
there are no anticorrelation in  Homestake data during the
years 1970-1977 (part of solar cycle \# 20). Also, the
latest data do not confirm the anticorrelation: large number
of the sunspots in 1991--1992 was accompanied by high
counting rate \cite{smirnov}. Therefore, one is tempted to
speculate
that, due to some reason, the anticorrelations are much more
prominent in the {\em odd-numbered\/}  solar cycles while
being
suppressed in the {\em even-numbered\/}  cycles. If we take
this
conjecture seriously, it would be easy to conclude that the
neutrino-depleting mechanism must somehow be correlated not
only with {\em the strength\/} of the solar magnetic field
but also with {\em the direction\/} of the toroidal solar
magnetic field which reverses every 11 years. Obviously,
this
feature would be difficult to accomodate within any of the
existing scenarios  except the present one.

\section{A SECOND NEUTRINO FLUX}

Apart from the reduction of the conventional
(i.e., thermonuclear) neutrino flux, a spectacular feature
of
our scenario is the prediction of a "second flux" of
electron neutrinos and antineutrinos from the Sun. While
thermonuclear neutrinos are produced due to the {\em weak\/}
interactions of the neutrino, the second flux arises due to
the {\em electromagnetic\/} production of neutrino-
antineutrino pairs. The most important process would be that
of plasmon decay into a neutrino-antineutrino pair. Thus the
second flux would consist of low-energy (about 200 {\rm
{}~eV})
 neutrinos produced in plasmon decays in the core of the
Sun, the number of such neutrinos being much greater than
that of the thermonuclear neutrinos. It would be
very interesting to consider the posibility of detecting
this second neutrino flux, about $ 10^{16}$ $~s^{-1}cm^{-2}$
in magnitude, on the Earth.

\section{CONCLUSION}

To conclude, in the context of the solar neutrino problem we
studied the consequences of the hypothesis that the electron
neutrino has a small but non-vanishing electric charge. The
main general consequence is that the solar neutrino flux can
be anisotropic. The cause of that anisotropy is the
antisymmetry of the large scale toroidal solar magnetic
field about the solar equatorial plane which leads to a
large gradient of the magnetic field along the direction
normal to the solar equatorial plane, denoted by $\langle
{\partial H_x \over \partial z} \rangle$. It is this
gradient that results in anisotropic deflection of the
neutrino flux. The magnitude of the anisotropy is controlled
by the product of the gradient, $\langle {\partial H_x \over
\partial z} \rangle$ and the neutrino charge, $\epsilon e$.
Arguments based on the energetics of the Sun show that the
neutrino charge must be less than $\epsilon \alt 10^{-13}$
\cite{brf}. Unfortunately, the value of the gradient
$\langle {\partial H_x \over \partial z} \rangle$ is not
very well known and, according to a rough estimate, may vary
from $10^{-7}$ G/cm to $10^{-
5}$ G/cm (or even perhaps up to $10^{-4}$ G/cm). We
calculated, in the linear approximation, the deficit of the
solar neutrino flux observed on the Earth caused by the
anisotropy of the neutrino flux at the surface of the Sun.
Assuming
that the neutrino charge is equal to $10^{-13}e$ and its
energy $E=0.8$ {\rm ~MeV} we
found the value of the gradient which is needed to obtain a
50\%  deficit by our mechanism: about $10^{-5}$ G/cm. If the
neutrino charge is much less than
$10^{-13}e$, it is unlikely to produce any
observable effect under the action of the magnetic field
{\em of the
convective zone\/}. (If one considers the magnetic field of
{\em the core and radiative zone\/}, it is not ruled out
that much less charges are still interesting, but any
quantitative conclusion on that point is difficult to reach
in view of our poor knowledge of those magnetic fields).

We then discussed some attractive experimental implications
of this scenario as well as the problems which have to be
solved so that this scenario could be considered as a full-
fledged solution to the solar neutrino puzzle.

We would like to stress that our mechanism has two
distinctive features which make it different from the other
proposed solutions (such as MSW, "just so", spin-flip, spin-
flavour precession and others):

1) Energy dependence of the effect: the less is the neutrino
energy, the more is the suppression of the neutrino flux;

2) 11-year variation: the neutrino flux is anticorrelated
with solar activity.\footnote{Note that such anticorrelation
may be absent if the deflection of the neutrino flux is
caused by the magnetic field of the solar core rather then
that of the convective zone.}

(In connection with these two features, we recall that:
spin-flip mechanism is energy {\em independent\/}; MSW
solution does {\em not\/} exhibit any 11-year variation.)

The second feature of our scenario is shared by all the
mechanisms employing the coupling of the neutrino magnetic
moment to the solar magnetic field in the convective zone
(involving or not involving the change of flavour). However,
our scenario is different from all those mechanisms in an
important way: the {\em seasonal variations \/} predicted by
those mechanisms are {\em not \/} predicted by our model.
The reason is that our effect is controlled by the {\em
gradient \/} of the magnetic field while all the others
depend on the magnetic field {\em itself\/}.

On the other hand, our model differs very clearly from the
MSW scenario or vacuum oscillations ("just so"). In these
oscillatory scenarios the missing electron neutrinos are
changed into neutrinos of other flavours ($\nu_{\mu}$ or
$\nu_{\tau}$). In our model, no $\nu_{\mu}$ or $\nu_{\tau}$
arise to replace the missing $\nu_{e}$. This difference will be
crucial for the forthcoming neutrino experiments involving
{\em neutral current\/} reactions (e.g., SNO or BOREXINO).
If an oscillatory scenario (MSW or "just so") is correct,
the number of observed neutral current events will be {\em
close \/} to the Standard Solar Model prediction (due to the
additional contribution of $\nu_{\mu}$ and $\nu_{\tau}$). On
the contrary, if our mechanism works, the observed number of
such events will be {\em less\/} than the theoretical
prediction.

But, independently of whether this scenario survives or not
in
its
present form, our arguments show that a more general problem
of the possible anisotropy of the neutrino flux due to the
interactions of the neutrino with the solar matter and
electromagnetic fields is certainly worth further pursuing.
Alternatively, the detailed observations of solar neutrinos
in future experiments can provide a unique way of obtaining
bounds on (or evidence of) the non-zero neutrino charge.

\section{acknowledgements}
The authors are grateful to N.Frankel, V.Gudkov, C.Horowitz,
A.Klein,
B.McKellar, E.Norman, A.Smirnov and R.Volkas for fruitful
discussions. A.I. is
indebted to A.Dymnikov for valuable help. This work was
supported in part by the Australian Research Council.

\end{document}